\documentclass[twocolumn,twoside,slac_two]{revtex4}
\usepackage{graphicx}
\usepackage{fancyhdr}
\usepackage{relsize}
\RequirePackage{xspace}
\def\babar   {\mbox{\slshape B\kern-0.1em{\smaller A}\kern-0.1em B\kern-0.1em{\smaller A\kern-0.2em R}}}
\def\peptwo  {PEP-II}
\providecommand{\Br}{\ensuremath{\mathcal{B}}\xspace}

\providecommand{\BDC}{\ensuremath{B^-\to D^{+}\pi^-\pi^-}\xspace}
\providecommand{\Bppp}{\ensuremath{B^+\to \pi^+\pi^{-}\pi^{+}}\xspace}
\providecommand{\Dppp}{\ensuremath{D_s^+\to \pi^{+}\pi^-\pi^+}\xspace}
\def\etalc              {{\it et~al.~}}   
\def\epem    {\ensuremath{e^+e^-}\xspace}
\def\ea{{\em et al.}}

\def\mev     {\ensuremath{\rm \,Me\kern -0.08em V}\xspace}
\def\Y       {\ensuremath{\Upsilon{\rm(4S)}}\xspace}

\def\CP                {\ensuremath{C\!P}\xspace}

\def\CalACP {{\ensuremath{{\cal A}_{\CP}}\xspace}}
\def\Abar {\kern 0.2em\overline{\kern -0.2em A}{}\xspace}
\def\Fbar {\kern 0.2em\overline{\kern -0.2em F}{}\xspace}
\def\eff     {\ensuremath{\epsilon}\xspace}
\def\cbar  {\ensuremath{\overline c}\xspace}

\def\Nev     {\ensuremath{N_\mathrm{event}}\xspace}
\def\Fbg     {\ensuremath{f_{\mathrm{bg}}}\xspace}

\def\MMminDP {\ensuremath{m^2_\mathrm{min}(D\pi)}\xspace}
\def\BR         {{\ensuremath{\cal B}\xspace}}

\def\Bbar    {\kern 0.18em\overline{\kern -0.18em B}{}\xspace}

\def\BB      {\ensuremath{B\Bbar}\xspace}
\def\Bz      {\ensuremath{B^0}\xspace}
\def\Bzb     {\ensuremath{\Bbar^0}\xspace}
\def\BzBzb   {\ensuremath{\Bz {\kern -0.16em \Bzb}}\xspace}
\def\Bu      {\ensuremath{B^+}\xspace}
\def\Bub     {\ensuremath{B^-}\xspace}
\def\Bp      {\ensuremath{\Bu}\xspace}
\def\Bm      {\ensuremath{\Bub}\xspace}
\def\Bpm     {\ensuremath{B^\pm}\xspace}

\def\BpBm    {\ensuremath{\Bu {\kern -0.16em \Bub}}\xspace}

\newcommand{\PPP}                {\mbox{$\pipm \pipm \pimp$}}
\newcommand{\fI}                 {\mbox{$\fz(980)$}}
\newcommand{\fz}                 {\mbox{$f_0$}}

\newcommand{\chiczpipm}          {\mbox{$\chiczero \pipm$}}

\newcommand{\chictwopipm}        {\mbox{$\chictwo \pipm$}}
\newcommand{\fIpipm}             {\mbox{$\fI \pipm$}}
\newcommand{\rhoII}              {\mbox{$\rhoz(1450)$}}

\newcommand{\rhoIIpipm}            {\mbox{$\rhoII \pipm$}}

\newcommand{\fII}                {\mbox{$f_2(1270)$}}

\newcommand{\fIIpipm}             {\mbox{$\fII \pipm$}}

\newcommand{\fIII}               {\mbox{$f_0(1370)$}}

\newcommand{\fIIIpipm}            {\mbox{$\fIII \pipm$}}

\def\piz   {\ensuremath{\pi^0}\xspace}

\def\pip   {\ensuremath{\pi^+}\xspace}
\def\pim   {\ensuremath{\pi^-}\xspace}
\def\pipi  {\ensuremath{\pi^+\pi^-}\xspace}
\def\pipm  {\ensuremath{\pi^\pm}\xspace}
\def\pimp  {\ensuremath{\pi^\mp}\xspace}

\newcommand{\rhoz}               {\mbox{$\rho^0$}}

\newcommand{\rhoI}               {\mbox{$\rhoz(770)$}}

\newcommand{\rhoIpipm}           {\mbox{$\rhoI \pipm$}}

\newcommand{\gevccSq}{\ensuremath{{\mathrm{\,Ge\kern -0.1em V^2\!/}c^4}}\xspace}
\newcommand{\gevccSqinv}{\ensuremath{{\mathrm{\,Ge\kern -0.1em V^{-2}}c^4}}\xspace}
                                                                                                                                        
\newcommand{\gevc}{\ensuremath{{\mathrm{\,Ge\kern -0.1em V\!/}c}}\xspace}
\newcommand{\mevc}{\ensuremath{{\mathrm{\,Me\kern -0.1em V\!/}c}}\xspace}
\newcommand{\gevcc}{\ensuremath{{\mathrm{\,Ge\kern -0.1em V\!/}c^2}}\xspace}
\newcommand{\mevcc}{\ensuremath{{\mathrm{\,Me\kern -0.1em V\!/}c^2}}\xspace}

\def\chiczero {\ensuremath{\chi_{c0}}\xspace}

\def\chictwo  {\ensuremath{\chi_{c2}}\xspace}
\def\MMmaxDP {\ensuremath{m^2_\mathrm{max}(D\pi)}\xspace}

\def\MMPP    {\ensuremath{m^2(\pi\pi)}\xspace}

\def\Dp      {\ensuremath{D^+}\xspace}
\def\pim     {\ensuremath{\pi^-}\xspace}

\def\Dtz     {\ensuremath{D^{*0}_2}\xspace}

\def\Dzz     {\ensuremath{D^{*0}_0}\xspace}

\def\KS    {\ensuremath{K^0_{\scriptscriptstyle S}}\xspace}

\begin{document}

\title{\boldmath Dalitz Plot Analyses of \BDC, \Bppp and \Dppp at \babar}


\author{Liaoyuan Dong\footnote{Email:~dongly@ihep.ac.cn}\footnote{Current Address: Institute of High Energy Physics, Beijing 100049, China} \\(On behalf of the \babar ~Collaboration)}
\affiliation{Iowa State University, Ames, Iowa 50011-3160, USA}

\begin{abstract}
We report on the Dalitz plot analyses of \BDC, \Bppp and \Dppp.
The Dalitz plot method and the most recent \babar ~results are discussed.
\end{abstract}

\maketitle

\thispagestyle{fancy}


\section{Introduction}
Dalitz plot analysis is an excellent way to study the dynamics of three-body decays.
The decays under study are expected to proceed predominantly through intermediate
quasi-two-body modes~\cite{twobody}
and experimentally this is the observed pattern. Dalitz plot analysis can explore resonances
properties, measure corresponding magnitudes and phases, and investigate corresponding \CP\ asymmetries.

We report on the Dalitz plot analyses of \BDC, \Bppp and \Dppp.
The data were collected with the \babar\ detector
at the \peptwo\ asymmetric-energy \epem\ storage rings at SLAC.
A detailed description of the \babar\ detector is given in Ref.~\cite{babarNim}.
Monte Carlo (MC) simulation is used to study the detector response, its acceptance, background,
and to validate the analysis.

\section{Dalitz Plot Analysis}
In the decay of a spin-0 particle $A$ to a final state composed of
three pseudo-scalar particles (abc), the differential decay rate is generally given in terms
of the Lorentz-invariant matrix element $\ensuremath{\mathcal{M}}$ by
\begin{eqnarray}
\label{dalitz}
\frac{d^2\Gamma}{dx dy}=\frac{| \ensuremath{\mathcal{M}} |^2}{256\pi^3m_A^3},
\end{eqnarray}
where $m_A$ is the $A$ particle mass, $x$ and $y$ are the invariant masses-squared of the $ab$ and $ac$ pairs, respectively.
The Dalitz plot provides a graphical representation of the variation of the square
of the matrix element, $|\ensuremath{\mathcal{M}}|^2$, over the kinematically
accessible phase space ($x$,$y$) of the process.

The distribution of candidate events in the Dalitz
plot is described in terms of a probability density function (PDF).
The PDF is the sum of signal and background components and has the form:
\begin{eqnarray}
\mathrm{PDF}(x,y) &=& \Fbg \frac{B(x,y)}{\int_{\mbox{DP}} B(x,y) dxdy} \nonumber \\
&+& (1-\Fbg) \frac{ \left [ S(x,y) \otimes {\cal{R}}\right ] \eff (x,y)}{ \int_{\mbox{DP}} \left [ S (x,y) \otimes {\cal{R}}\right ] \eff (x,y) dxdy}, \nonumber \\
\label{pdfall}
\end{eqnarray}
where
\begin{itemize}
\item The integral is performed over the whole Dalitz plot (DP).
\item $S$ ($=|\ensuremath{\mathcal{M}}|^2$) and $B$ describe signal and background amplitude contributions, respectively.
\item The $S(x,y) \otimes {\cal{R}}$ is the signal term convolved with the signal resolution function,
      the signal resolution can be safely neglected for broad resonances.
\item \Fbg is the fraction of background events.
\item \eff is the reconstruction efficiency.
\end{itemize}

An unbinned maximum likelihood fit to the Dalitz plot is performed in order to maximize the value of
\begin{eqnarray}
{\cal L}= \prod_{i=1}^{\Nev} \mathrm{PDF}(x_i,y_i)
\label{eq:likelihood}
\end{eqnarray}
with respect to the parameters used to describe $S$,
where $x_i$ and $y_i$ are the values of $x$ and $y$ for event $i$ respectively,
and \Nev is the number of events in the Dalitz plot.
In practice, the negative-log-likelihood (NLL) value
\begin{eqnarray}
\mathrm{NLL}=-\ln{\cal L}
\end{eqnarray}
is minimized in the fit.

The isobar model formulation is used, in which the signal decays are described by a coherent sum of a number of two-body amplitudes.
The total decay matrix element $\ensuremath{\mathcal{M}}$ is given by
\begin{eqnarray}
\ensuremath{\mathcal{M}} = \sum_k \rho_{k}e^{i\phi_{k}} A_k,
\end{eqnarray}
where $\rho_{k}$ and $\phi_{k}$ are the magnitudes and phases of the $k^{\rm th}$ decay mode respectively,
the $A_k$ distributions for resonant contribution describe the dynamics of the decay amplitudes and are written as the product of a
relativistic Breit-Wigner function, two Blatt-Weisskopf barrier form factors~\cite{formf} and an angular function~\cite{Zemach}.

The efficiency-corrected fraction due to the resonant or non-resonant contribution $k$ is defined as follows:
\begin{eqnarray}
f_k = \frac {|\rho_k|^2 \int_{\mbox{DP}} |A_k|^2 dx dy}
{\int_{\mbox{DP}} \mid \mathcal{M} \mid ^2 dx dy}.
\end{eqnarray}
The $f_k$ values do not necessarily add to 1 because of interference effects.
The uncertainty on each $f_k$ is evaluated by propagating the
full covariance matrix obtained from the fit.

\section{Results from \babar}

\subsection{\BDC~\cite{bdpp}}

\begin{figure}[h]
\centering
\includegraphics[width=75mm]{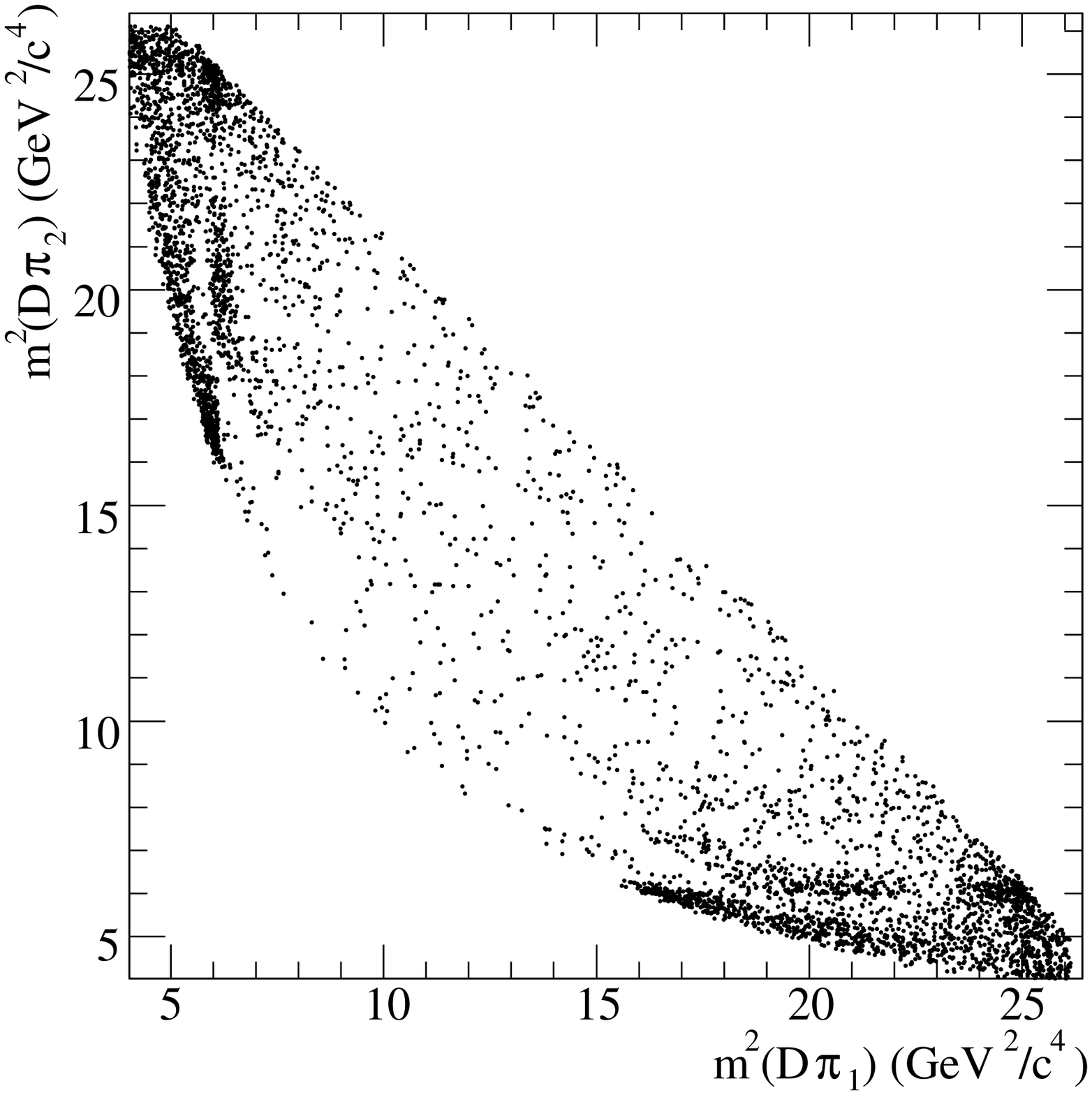}
\caption{Dalitz plot for \BDC.} 
\label{fig:datadalitz}
\end{figure}

We reconstruct the decays \BDC~\cite{conjugate} with $D^+\to K^-\pi^+\pi^+$ based on a sample of about
$383\times 10^6$ $\Y \to \BB$ decays.
The resulting Dalitz plot distribution for data is shown in Fig.~\ref{fig:datadalitz}.
The number of signal events in the Dalitz plot is $3414\pm85$.
The background fraction is about 30.4\%.
The background is parametrized using the MC simulation.

The nominal fit model contains 
$\Dtz$, $\Dzz$, $D^*(2007)^0$, {\ensuremath{B^{*}}} and $\mathcal{P}$-wave non-resonant contributions.
The $\mathcal{S}$-wave and $\mathcal{D}$-wave non-resonant contributions are found to be small and can be ignored.
Replacing the $D^*(2007)^0$ by a $D\pi$ $\mathcal{S}$-wave~\cite{dpi} state is found to give a much worse fit.

\begin{figure}[h]
\centering
\includegraphics[width=80mm]{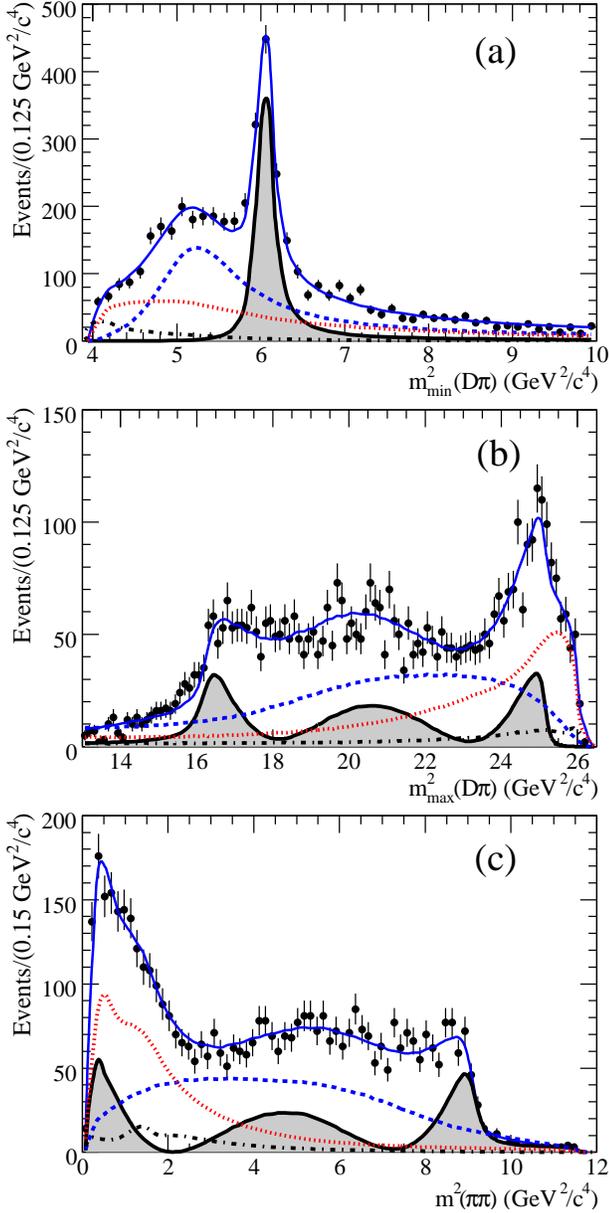}
\caption{Result of the nominal fit to the data: projections on 
(a) \MMminDP, (b) \MMmaxDP  and (c) \MMPP.
The points with error bars are data, the solid curves represent the nominal fit.
The shaded areas show the \Dtz contribution,
the dashed curves show the \Dzz signal, the dash-dotted curves show the $D^*(2007)^0$
and {\ensuremath{B^{*}}} signals, and the dotted curves show the background.}
\label{fig:datafit1}
\end{figure}

Fig.~\ref{fig:datafit1}a, \ref{fig:datafit1}b and \ref{fig:datafit1}c show the
\MMminDP, \MMmaxDP and \MMPP projections respectively.
Here $\MMmaxDP$ and $\MMminDP$ are the heavy and light invariant masses-squared
of the $D\pi$ systems, respectively.
The distributions in Fig.~\ref{fig:datafit1} show good agreement between the data and the fit.

We have performed some tests with the broad resonance \Dzz excluded or with the
$J^P$ of the broad resonance replaced by other quantum numbers.
In all cases, the NLL values are significantly worse than that of the nominal fit.
We conclude that a broad spin-0 state \Dzz is required in the fit to the data.

The total branching fraction of the \BDC decay is measured to be
\begin{eqnarray}
\Br(\BDC) = (1.08 \pm 0.03 \pm 0.05) \times 10^{-3}, \nonumber
\end{eqnarray}
where the first error is statistical and the second is
systematic.
The systematic error on the measurement of the total \BDC branching fraction
is due to the uncertainties on 
the number of \BpBm events,
the charged track reconstruction and identification efficiencies,
the $\Dp \to K^-\pi^+\pi^+$ branching fraction,
the background shape and the fit models.

The mass and width of \Dtz are determined to be:
\begin{eqnarray}
m_{\Dtz} &=& (2460.4\pm1.2\pm1.2\pm1.9) \mevcc  \mbox{~and} \nonumber\\
\Gamma_{\Dtz} &=& (41.8\pm2.5\pm2.1\pm2.0) \mev, \nonumber
\end{eqnarray}
respectively, while for the \Dzz they are:
\begin{eqnarray}
m_{\Dzz} &=& (2297\pm8\pm5\pm19) \mevcc \mbox{~and} \nonumber\\
\Gamma_{\Dzz} &=& (273\pm12\pm17\pm45) \mev, \nonumber
\end{eqnarray}
where the first and second errors reflect the statistical    
and systematic uncertainties, respectively, the third one
is the uncertainty related to the fit models
and the Blatt-Weisskopf barrier factors.

We have also obtained exclusive branching fractions for $D^{*0}_2$ and $D^{*0}_0$ production:
\begin{eqnarray}
\Br(B^- &\to& D^{*0}_2 \pi^-) \times \Br(D^{*0}_2 \to D^+\pi^-) \nonumber\\
&=& (3.5\pm0.2\pm0.2\pm0.4) \times 10^{-4} \mbox{~and} \nonumber\\
\Br(B^- &\to& D^{*0}_0 \pi^-) \times \Br(D^{*0}_0 \to D^+\pi^-) \nonumber\\
&=& (6.8\pm0.3\pm0.4\pm2.0) \times 10^{-4}. \nonumber
\end{eqnarray}

The relative phase of the scalar and tensor amplitude is measured to be
\begin{eqnarray}
\phi_{D^{*0}_0} = -2.07\pm0.06\pm0.09\pm0.18 \mbox{~rad}. \nonumber
\end{eqnarray}

The systematic uncertainties that affect the results of Dalitz plot analysis
are due to the background parametrization, the selection criteria,
and the possible fit bias.

Our results for the masses, widths and branching fractions are consistent
with but more precise than previous measurements performed by Belle~\cite{belle-prd}.

\subsection{\Bppp~\cite{bppp}}

We fit the $B^-$ and $B^+$ samples independently in order to determine the
\CP\ asymmetry. The total signal amplitudes for \Bp\ and \Bm\ decays are given by
\begin{eqnarray}
\label{eq:totAmp}
  \mathcal{M} &=&
  \sum_j c_j A_j(m_{\rm max}^2,m_{\rm min}^2) \,,\nonumber \\
  \bar{\mathcal{M}} &=&
  \sum_j \cbar_j \bar{A}_j(m_{\rm max}^2,m_{\rm min}^2) \,,
\end{eqnarray}
where $m_{\rm max}^2$ and $m_{\rm min}^2$ are the heavy and light invariant masses-squared of
the $\pi^{\pm}\pi^{\mp}$ systems, respectively.
The complex coefficients $c_j$ and $\cbar_j$ for a given decay mode $j$
contain all the weak phase dependence. Since the $A_j$ terms contain only strong dynamics,
$A_j \equiv \bar{A}_j$. 
The \CP\ asymmetry for each contributing resonance is determined from the fitted parameters
\begin{eqnarray}
  \label{eq:cpasym}
  \CalACP_{\!,\,j} & = &
  \frac
  {\left|\cbar_j\right|^2 - \left|c_j\right|^2}
  {\left|\cbar_j\right|^2 + \left|c_j\right|^2}\, .
\end{eqnarray}

\begin{figure}[h]
\centering
\includegraphics[width=80mm]{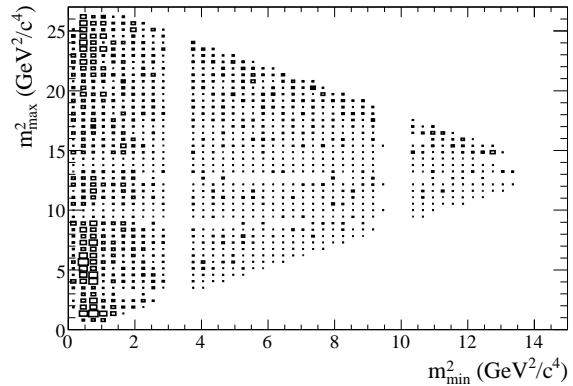}
\caption{Background-subtracted Dalitz plot for \Bppp. The plot shows bins with greater than zero
      entries. The depleted bands are the charm and charmonia exclusion regions.}   
\label{fig:Bpppdalitz}
\end{figure}

We reconstruct the decays \Bppp based on a sample of about
$465\times 10^6$ $\Y \to \BB$ decays.
We reject background from two-body decays of $D^0$ meson and charmonium
states ($J/\psi$ and $\psi(2S)$) by excluding invariant masses (in units of
$\rm{Ge\kern -0.1em V\!/}c^2$) in the ranges:
$1.660 < m_{\pipi} < 1.920$,
$3.051 < m_{\pipi} < 3.222$, and
$3.660 < m_{\pipi} < 3.820$.
The background-subtracted Dalitz plot distribution for data is shown in Fig.~\ref{fig:Bpppdalitz}.
We fit $4335$ $B$ candidates to calculate the fit fractions and \CP\ asymmetries. 
The background is parametrized using the MC simulation.

The nominal signal Dalitz plot model comprises a momentum-dependent
non-resonant component and four intermediate resonance states:
\rhoIpipm, \rhoIIpipm, \fIIpipm, and \fIIIpipm.
We do not find any significant contributions from 
$f_0(980)\pi^\pm$, $\chi_{c0}\pi^\pm$, or $\chi_{c2}\pi^\pm$.
The absence of the charmonium contributions precludes the extraction of
the unitarity triangle angle $\gamma$.
We find no evidence for direct \CP\ violation.
The inclusive \CP\ asymmetry is determined to be $\left(+3.2\pm 4.4\pm 3.1\,^{+2.5}_{-2.0}\right)\%$, where the
uncertainties are statistical, systematic, and model-dependent, respectively.

\begin{figure}[thb]
  \begin{center}
    \includegraphics[width=0.98\columnwidth]{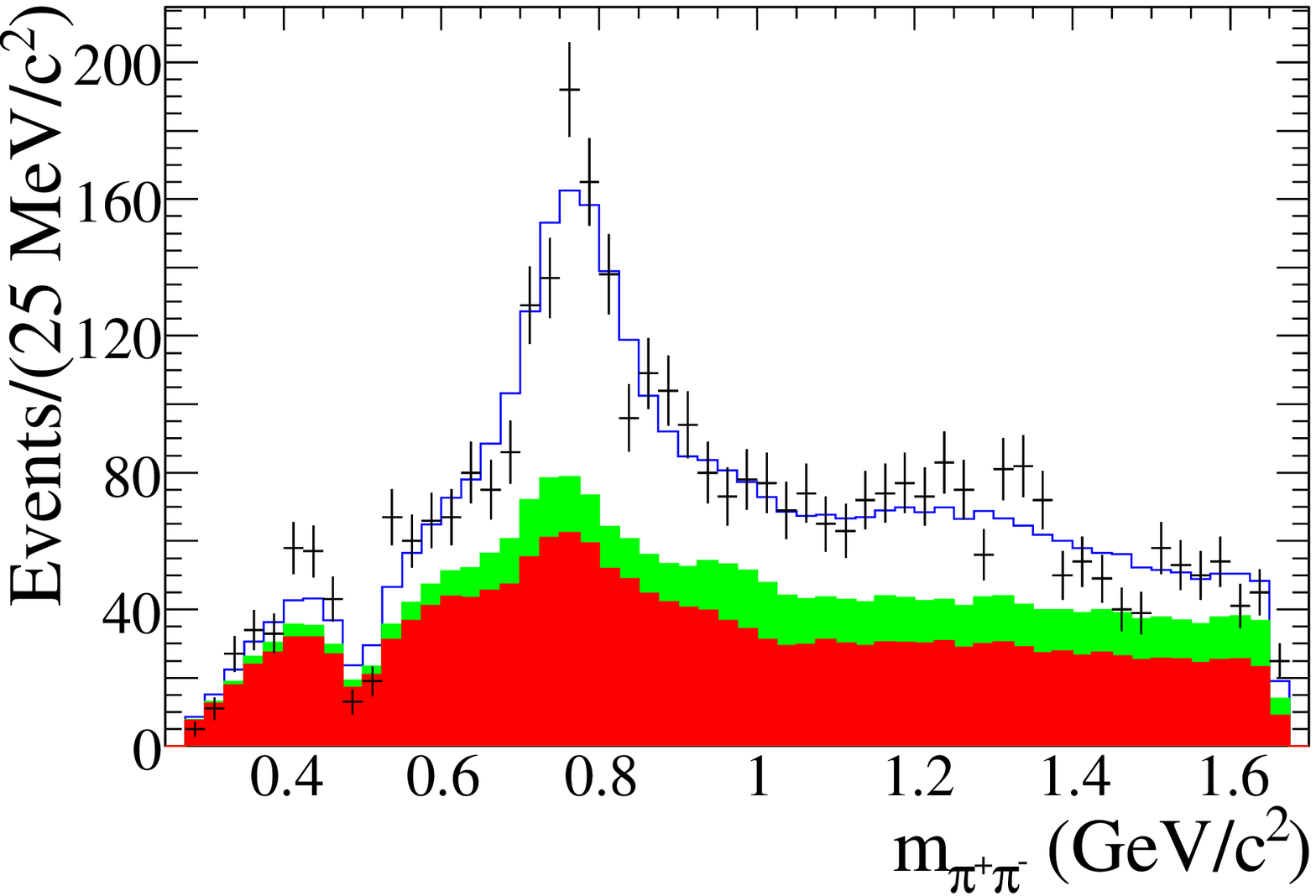}
    \includegraphics[width=0.98\columnwidth]{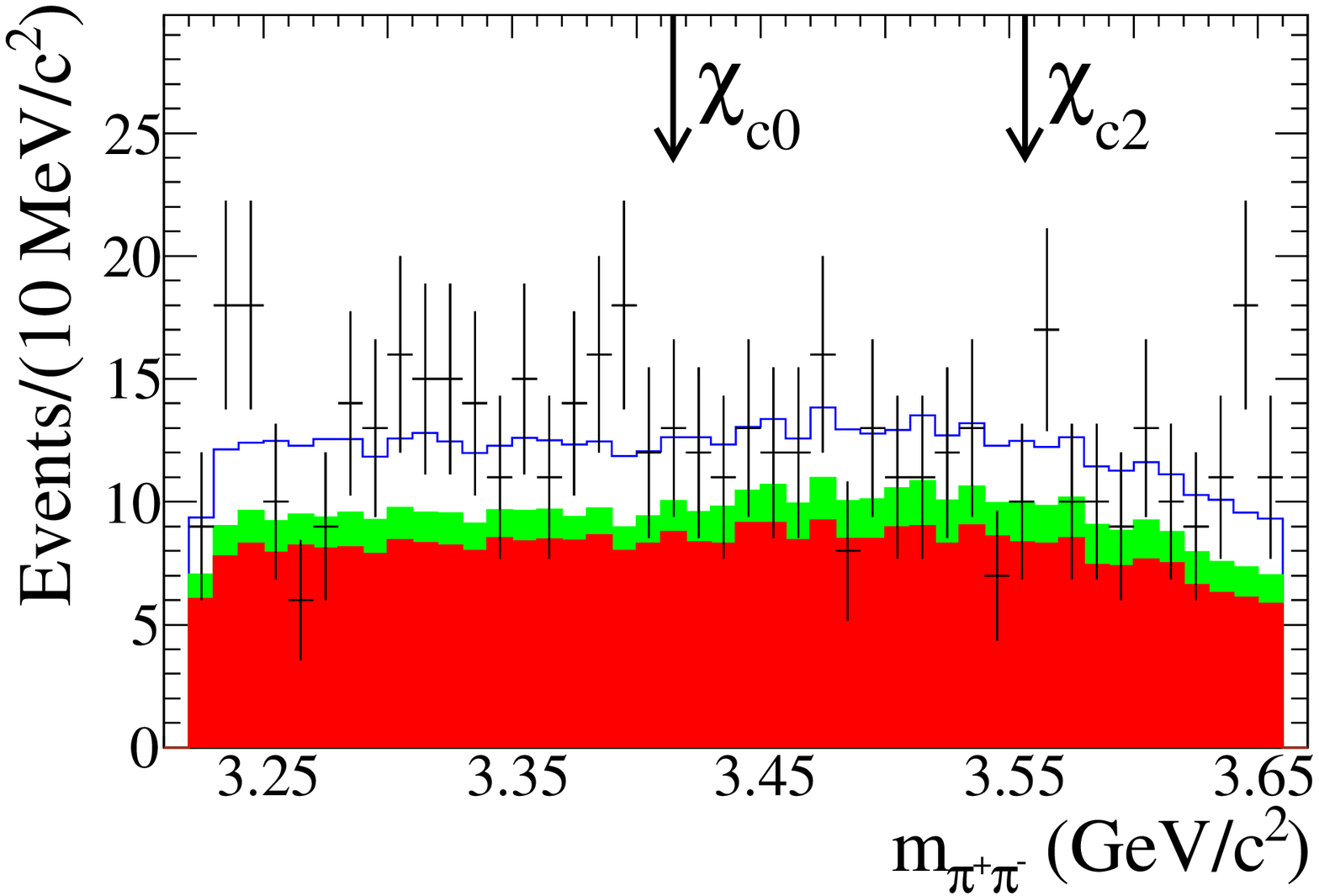}
    \caption{
      Dipion invariant mass projections: 
      (upper) in the \rhoI\ region; and
      (lower) in the regions of $\chi_{c0}$ and $\chi_{c2}$.
      The data are the points with statistical error bars,
      the dark-shaded (red) histogram is the $q\bar{q}$ component, 
      the light-shaded (green) histogram is the \BB\ background
      contribution, while the upper (blue) histogram shows the total fit result. 
      The dip near 0.5\gevcc in the upper plot is due to the rejection of
      events containing \KS\ candidates.
    }
    \label{fig:nom}
  \end{center}
\end{figure}

Projections of the data, with the fit result overlaid, as $\pipm\pimp$
invariant-mass distributions can be seen in Fig.\ref{fig:nom}.
The agreement between the fit results and the data is generally good.

The Dalitz plot is dominated by the $\rhoI$ resonance and
a non-resonant contribution.
The mass and width of the \fIII\ are determined to be 
$m_{f_0(1370)} = 1400\pm40\mevcc$ and $\Gamma_{f_0(1370)} = 300\pm80\mev$ (statistical
uncertainties only).
We measure the branching fractions 
$\BR(\Bpm \to \PPP) = (15.2\pm0.6\pm1.2\pm0.4)\times 10^{-6}$,
$\BR(\Bpm \to \rhoIpipm) = (8.1\pm0.7\pm1.2^{+0.4}_{-1.1})\times 10^{-6}$,
$\BR(\Bpm \to \fIIpipm) = (1.57\pm0.42\pm0.16\,^{+0.53}_{-0.19})\times 10^{-6}$,
and
$\BR(\Bpm \to \PPP\ {\rm{non-resonant}}) = (5.3\pm0.7\pm0.6^{+1.1}_{-0.5})\times 10^{-6}$,
where the uncertainties are statistical, systematic, and model-dependent, respectively.
We have made the first observation of the decay $\Bpm\to\fIIpipm$
with a statistical significance of $6.1\sigma$.

The systematic uncertainties that affect the results of Dalitz plot analysis
are due to the the number of \BpBm events,
signal and background PDF parametrization,
the charged track reconstruction and identification efficiencies,
and the possible fit bias.

Our results will be useful to reduce model uncertainties in the extraction of
the CKM angle $\alpha$ from time-dependent Dalitz plot analysis of
$\Bz\to\pip\pim\piz$.

\subsection{\Dppp~\cite{dppp}}

\begin{figure}[h]
\centering
\includegraphics[width=80mm]{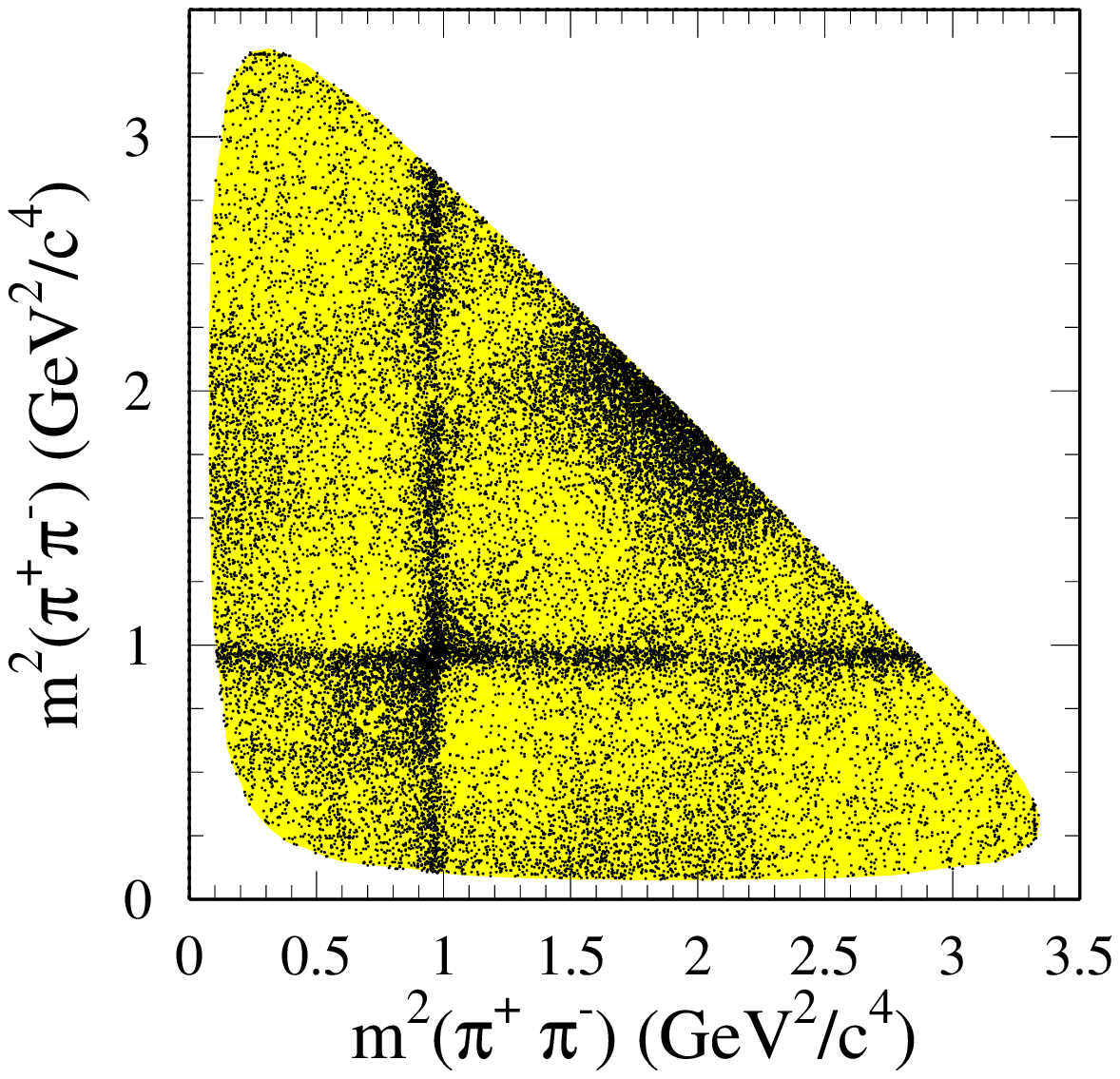}
\caption{Dalitz plot for \Dppp.}
\label{fig:dpppdalitz}
\end{figure}

We reconstruct $D^+_s \to \pip \pim \pip$ based on 384~${\rm fb}^{-1}$ data sample.
We use the tag of $D^*_s(2112)^+ \to D^+_s \gamma$ to reduce the combinatorial background.
The signal PDFs are build using the $D^+_s \to K^+ K^- \pip$.
The reconstructed $D^+_s$ sample contains 13179 events with a purity of 80\%.
The background shape is obtained by fitting the $D^+_s$ sidebands.
The Dalitz plot distribution for data is shown in Fig.~\ref{fig:dpppdalitz}.

For the $\pip \pim$ $\mathcal{S}$-wave amplitude, we use a 
model-independent partial wave analysis: instead of
including the $\mathcal{S}$-wave amplitude as a superposition of relativistic Breit-Wigner functions, we divide
the $\pip\pim$ mass spectrum into 29 slices and we parametrize the $\mathcal{S}$-wave by an
interpolation between the 30 endpoints in the complex plane: 
\begin{eqnarray}
A_{\mathcal{S}-{\rm wave}}(m_{\pi\pi}) = {\rm Interp}(c_k(m_{\pi\pi})e^{i\phi_k(m_{\pi\pi})})_{k=1,..,30}\,. \nonumber \\
\end{eqnarray}
The amplitude and phase of each endpoint are free parameters. The width of each slice is tuned to get
approximately the same number of $\pip \pim$ combinations ($\simeq 13179 \times 2/29$).
Interpolation is implemented by a Relaxed Cubic Spline~\cite{cern1}. The phase is not constrained
in a specific range in order to allow the spline to be a continuous function.

\begin{figure*}[t]
\begin{center}
\includegraphics[width=10cm]{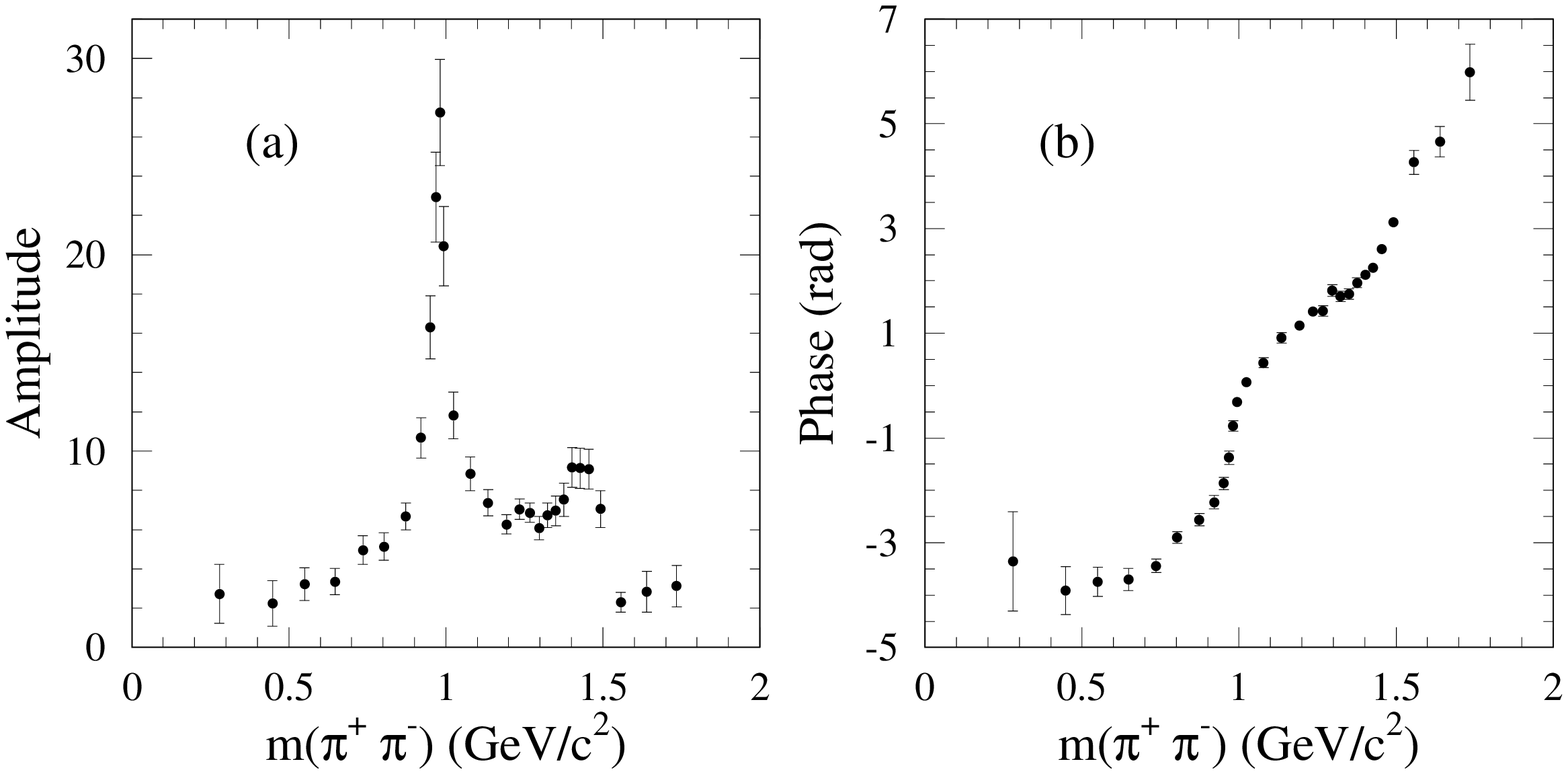}
\caption{(a) $\mathcal{S}$-wave amplitude extracted from the best fit, (b) corresponding $\mathcal{S}$-wave phase.}
\label{fig:fig_2}
\end{center}
\begin{center}
\includegraphics[width=10cm]{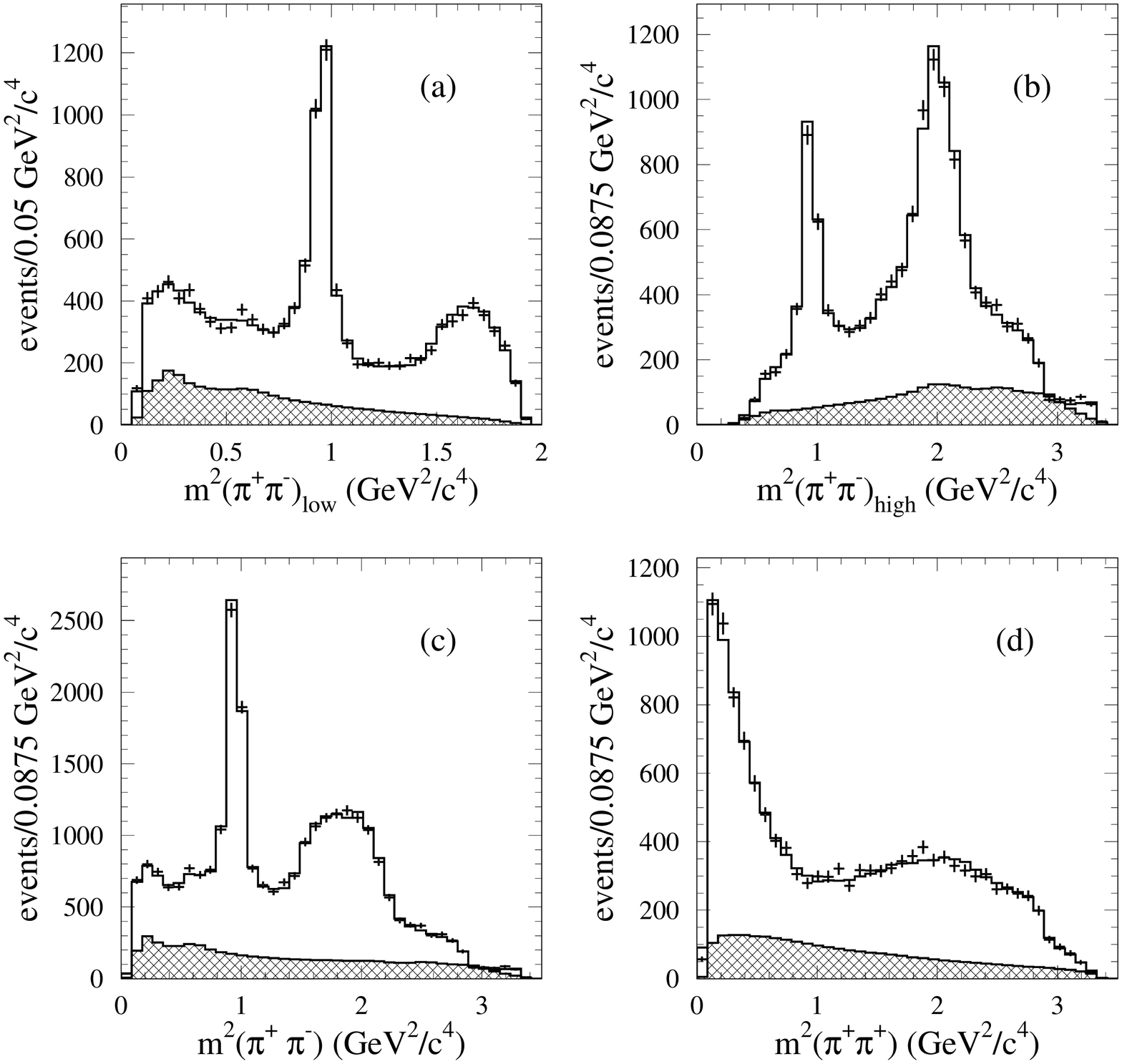}
\caption{Dalitz plot projections (points with error bars) and fit results (solid histogram).
(a) $m^2(\pip \pim)_{{\rm low}}$, (b) $m^2(\pip \pim)_{{\rm high}}$, (c) total $m^2(\pip \pim)$, (d) $m^2(\pip \pip)$.
The hatched histograms show the background distribution.}
\label{fig:fig_3}
\end{center}
\end{figure*}

The nominal signal Dalitz plot model includes $f_2(1270)$, $\rho(770)$, $\rho(1450)$ and $\pi\pi$ $\mathcal{S}$-wave.
The Dalitz plot is dominated by the $D_s \to (\pip \pim)_{\mathcal{S}-{\rm wave}} \pip$ contribution.
The $\mathcal{S}$-wave shows, in both amplitude and phase, the expected behavior for the $f_0(980)$ resonance,
and further activity in the regions of the $f_0(1370)$ and $f_0(1500)$ resonances.
The $\mathcal{S}$-wave is small in the $f_0(600)$ region, indicating that this resonance has a small coupling to $s \bar s$.

The resulting $\mathcal{S}$-wave $\pip \pim$ amplitude and phase is shown in
Fig.~\ref{fig:fig_2}a and \ref{fig:fig_2}b.
Our results on the amplitude and phase of the
$\pip \pim$ $\mathcal{S}$-wave agree better (within uncertainties) with the results from FOCUS~\cite{focus} than those from E791~\cite{e791_ds}.
The Dalitz plot projections together with the fit
results are shown in Fig.~\ref{fig:fig_3}.

Since $D_s^+ \to \pip \pim \pip$ and $D_s^+ \to K^+ K^- \pip$ have similar topologies,
the ratio of branching fractions is expected to have a reduced
systematic uncertainty. We therefore select events from the two $D_s$ decay modes using similar selection criteria
for the $D^{*+}_s$ selection.
The ratio of branching fractions is evaluated as:
\begin{eqnarray}
\frac{ \BR(D_s^+ \to \pip \pim \pip)}{ \BR(D_s^+ \to K^+ K^- \pip)} = 0.199 \pm 0.004 \pm 0.009 , \nonumber
\end{eqnarray}
where the first error is statistical and the second is
systematic.
The systematic uncertainty comes from the MC statistics and from
the selection criteria used.

\section{Conclusion}
We reviewed the recent results from the Dalitz plot analyses of \BDC, \Bppp and \Dppp 
performed by the \babar~ Collaboration. 
The results can be summarized as follows:
\begin{itemize}
\item \BDC \\
We find the total branching fraction of the three-body decay: 
$\Br(\BDC) = (1.08 \pm 0.03\pm 0.05) \times 10^{-3}$.
We observe the established \Dtz and confirm the existence of \Dzz
in their decays to $D^+\pi^-$,
we measure the masses and widths of \Dtz and \Dzz to be:
$m_{\Dtz} = (2460.4\pm1.2\pm1.2\pm1.9) \mevcc$,
$\Gamma_{\Dtz} = (41.8\pm2.5\pm2.1\pm2.0) \mev$,
$m_{\Dzz} = (2297\pm8\pm5\pm19) \mevcc$ and $\Gamma_{\Dzz} = (273\pm12\pm17\pm45) \mev$.  

\item \Bppp \\
We measure the branching fractions 
$\BR(\Bpm \to \PPP) = (15.2\pm0.6\pm1.2\pm0.4)\times 10^{-6}$,
$\BR(\Bpm \to \rhoIpipm) = (8.1\pm0.7\pm1.2^{+0.4}_{-1.1})\times 10^{-6}$,
$\BR(\Bpm \to \fIIpipm) = (1.57\pm0.42\pm0.16\,^{+0.53}_{-0.19})\times 10^{-6}$, and
$\BR(\Bpm \to \PPP\ {\rm{non-resonant}}) = (5.3\pm0.7\pm0.6^{+1.1}_{-0.5})\times 10^{-6}$.
We observe no significant direct \CP\ asymmetries for the above modes, and
there is no evidence for the decays $\Bpm\to\fIpipm$, $\Bpm\to\chiczpipm$, or $\Bpm\to\chictwopipm$.

\item \Dppp \\
We perform a high precision measurement of the ratio of branching fractions:
$\BR(D_s \to \pip \pim \pip)/\BR(D_s \to K^+ K^- \pip)=0.199 \pm 0.004
\pm 0.009$. 
The amplitude and phase of the
$\pip \pim$ $\mathcal{S}$-wave is extracted in a model-independent way for the first time.

\end{itemize}

\bigskip 

\begin{thebibliography}{9}
\bibitem{conjugate} Charged conjugate states are implied throughout the paper.
\end{thebibliography}

\end{document}